\documentstyle[12pt,titlepage,fleqn]{article}

\newcommand{\be}{\begin{equation}}
\newcommand{\ee}{\end{equation}}
\newcommand{\ba}{\begin{eqnarray}}
\newcommand{\ea}{\end{eqnarray}}

\newcommand{\cO}{{\cal O}}
\newcommand{\non}{\nonumber}

\newcommand{\oh}{\frac{1}{2}}

\begin{document}

\begin{titlepage}

\hfill IC/95/57

\vspace*{0.1truecm}
\begin{center}
\vskip 1.0cm
{\large\bf
$(0,2)$ STRING COMPACTIFICATIONS:\\
THE HIGGS MECHANISM }

\vskip 2.0cm

{\large Fermin ALDABE\footnote{E-mail: aldabe@ictp.trieste.it}}

{\large\em International Center for Theoretical Physics\\
P.O. Box 586, 34100 Trieste, Italy}\\
\vspace{.5in}
\today \\
\vspace{.5in}
{\bf ABSTRACT}\\
\begin{quotation}
\noindent
We present  the Higgs mechanism in (0,2) compactifications.
 The existence of a vector bundle data duality (VBDD)
 in $(0,2)$ compactifications which
is present at the Landau-Ginzburg point allows us to connect in a
smooth manner theories with different gauge groups with the same base
manifold and same number of effective generations.
As we move along the Kahler moduli space of the theories
with $E_6$ gauge group, some of the gauginos pick up
masses and break the gauge group to $SO(10)$ or $SU(5)$.

\end{quotation}
\end{center}

\end{titlepage}
\section{Introduction}
The heterotic string is believed to be the correct description of particle
physics.
A Part of this theory is
compactified to a three complex dimension Kahler manifold with
vanishing Ricci tensor, a Calabi-Yau manifold,
and much effort has been invested in understanding how
this part of the theory affects the remaining four dimensional theory
which should describe the low energy effective field theory.  In particular,
the internal manifold will determine the number of generations of the
low energy effective field theory, while the choice of the complex structure
deformations and Kahler structure deformations determine the Yukawa
couplings.  Thus a thorough understanding of the internal theory is needed
to relate the possible vacuum configurations of this theory, which is
determined by the
structure of the manifold.  In \cite{bat}
\cite{w1} \cite{agm}, $(2,2)$ strings propagating on
birationally distinct manifolds were related with the use
of mirror symmetry \cite{gp}.
Of phenomenological importance are the $(0,2)$ compactifications  because
they admit gauge groups of smaller rank than $E_6$, the only gauge group
admitted by $(2,2)$ compactifications.  Recently \cite{d2}, it has been
shown that
two $(0,2)$ models propagating on different pairs $(M,E)$, $M$ being the base
manifold and $E$ the vector bundle, having different Euler characters for
the base manifolds
are dual.  The duality transformation exchanges the gauge group moduli with the
gravitational moduli.
This
duality occurs for both models when the Kahler class is negative, or
equivalently,  at the Landau-Ginzburg phase.

Here we
present a different duality where two pairs $(M,E)$ which have  same
base manifold $M$  but different vector bundles,
are dual theories.  It is shown that this duality
takes place at the Landau-Ginzburg point.  The rank of the gauge groups
of the two pairs will be different.  One will have an $E_6$ space time
gauge group
while the other will have an $SO(10)$ space time gauge group.  In order
to achieve this duality one of the models will have to exhibit a Higgs
mechanism as we shall see.

This article is organized as follows.  Section two reviews the Calabi-Yau
Landau-Ginzburg \cite{vw}
correspondence presented in \cite{w1} for $(2,2)$ models.
It is in this section
that we show that in the Calabi-Yau region where instantons effects
are small, the description of the model needs the information of both the
hypersurface and the tangent bundle.
The information they carry
are equivalent.  In the
Landau-Ginzburg region, where the instanton effects are large, and the
perturbative treatment of them is not possible,
 we show that the information of the hypersurface is lost and the only data
that remains is that of the tangent bundle.
In section three we review the $(0,2)$ gauged linear $\sigma$   models,
deformations \cite{w1} of (2,2) models and as well as the Higgs (0,2)
model.
In section four, we present examples of models in which the gauge
group $E_6$ is broken to $SO(10)$ with out changing the number of net
generations.  In section 5 we show how Calabi-Yau manifolds with
$E_6$ are connected in a contunous manner
to the same Calabi-Yau manifold with
$SU(5)$ gauge group
preserving the number of effective generations.  The last section
has some conclusions.

\section{Calabi-Yau/Landau-Ginzburg Phases of $(2,2)$ Models}

As pointed out \cite{w1}, it is possible to relate a string propagating on
a Calabi-Yau manifold to Landau-Ginzburg orbifolds by means of a linear
$\sigma$
model.  We review this relation, by following the  construction in \cite{w1}.

The $N=2$ supersymmetric linear $\sigma$ model in two dimensions
is constructed by dimensionally reduced N=1 supersymmetry in four
dimensions.   These models contain the following fields.
One $U(1)$
gauge vector multiplet $ A$ whose bosonic vector has field components
$a_1,a_2$.
Of importance will be the presence of the auxiliary field $D$ in this
multiplet.
There will also be chiral superfields $X_i$, $i=1,...,n$
which transform in the fundamental of $U(1)$ and which
are charged
with respect to the $U(1)$ gauge field.  Their bosonic
component will be denoted by  $x_i$. Finally, there will
be other chiral superfields $P_j$
 which are charged  and also transform
in the fundamental of $U(1)$,
its bosonic component we will denote by $p_j$.

The Lagrangian for the liner $\sigma$ model
in terms of these superfields will be \cite{w1}
\be
L=L_{kin}+L_{W}+L_{gauge}+L_{D,\theta}
\ee
with
\ba
L_{kin}&=&\int d^2y d^4\theta\sum_i\bar{X_i}e^{2Q_i  A}X_i
+\bar{P_j}e^{2 Q_j  A}P_j\non\\
L_{W}&=&
-\int d^2y d^2\theta P_j W^j(X)+h.c.\non\\
L_{gauge}&=&-\frac{1}{4}\int d^y d^4\theta \bar{\cal A}{\cal A}\non\\
L_{D,\theta}&=&\frac{it}{2\sqrt{2}}\int d^2y  d\theta^+d\bar{\theta}^-
\bar{\cal A}|_{\theta^-=\bar{\theta}^+=0}-
\frac{i\bar{t}}{2\sqrt{2}}\int d^2y  d\theta^-d\bar{\theta}^+
\bar{\cal A}|_{\theta^+=\bar{\theta}^-=0}\non\\
\label{potu}
\ea
Here, ${\cal A}$ is the field strength of the supergauge field.
The potentials $W_j$, $j=1,...,n-4$, will be homogeneous transverse
polynomials of degree $d_j$
in the fields $X_i$.
$t=\theta+ir$ will turn out to be a good coordinate over the moduli space of
Khaler structure deformations.
Performing the superspace integration and integrating out the auxiliary
fields of the chiral and gauge superfields
we obtain the following bosonic potential
\ba
U&=&\sum_j|W_j(x)|^2+\sum_i|\sum_jp_j\frac{\partial W_j}{\partial x_i}|^2
+|\sigma|^2(|\sum_i|q_i|^2|x_i|^2+|q_j|^2|p_j|^2)\non\\
&&+
\oh (|\sum_iq_i|x_i|^2+q_j|p_j|^2-r)^2\label{U}
\ea
where $\sigma$ is the scalar component of the $A$ vector multiplet.
The last term in $U$ must vanish because it is proportional to
$D^2$, since only if it vanishes will supersymmetry be preserved.
If we consider the situation in which the charges $q_i$ are all positive,
the $q_j$'s are all negative, and $\sum_jq_j=-\sum_i q_i$ then when
$r>>1$ all $x_i$ cannot vanish
and since we have chosen $W_j$'s to be transverse it follows that the
$p_j$'s must
vanish. Also $a_1$ and $a_2$ pick a mass of order $\sqrt{r}$ and
remain excluded from the massless spectrum. The last term
in (\ref{U}) will then describe a manifold which  after
moding out by the $U(1)$ symmetry induced by the gauge field,
will be a weighted complex projective space of complex dimension $n-1$,
${\bf WCP}^{n-1}_{q_1, ...,q_n}$
with Kahler class proportional to $r$.
The vanishing of the first term in (\ref{U}) will define a hypersurface,
given by the locii of $W_j(x)$, embedded in ${\bf WCP}^{n-1}_{q_1,...,q_n}$.
We will call
this manifold $M$.  Since the sum of the degrees of the polynomials $W_j$  has
been
chosen to be equal to the sum of the weights of ${\bf
WCP}_{q_1,...,q_n}^{n-1}$,
the first Chern class will vanish implying
that $M$ is Ricci flat, a condition needed to allow for a conformal
field theory on a compact manifold with complex dimension $3$.
This way we can construct a string propagating on a
Calabi Yau manifold starting from a linear $\sigma$ model.

Let us analyze the case in which $r<<-1$.  Inspection of (\ref{U})
subject to the condition that supersymmetry be preserved
shows that the $p_j$'s
cannot vanish simultaneously.
By transversality of $W_j$ we conclude that
all $x_i$ must vanish and that $a_1$ and $a_2$ pick up a mass of order
$\sqrt{r}$ and drop out of the massless spectrum once more.  The fields
$p_j$ describe a ${\bf WCP}_{\{q_j\}}^{n-5}$ whose Kahler class
is of order $\cO(\sqrt{r})$.
We may now expand
about the classical solutions.  We find that the fields $x_i$ remain massless
for our particular case and that we may integrate out $p_j$ by setting it to
its expectation value.
What remains after rescaling some of the fields is shown below
\be
U=\sum_i|\sum_j\frac{\partial W^j}{\partial x_i}|^2
\label{Ulg}.
\ee
or in superspace coordinates
\be
L_{W_eff}= -\int d^2y d^2\theta W^j(X)
\ee
which is a Hibrid Landau-Ginzburg (HLG) superpotential which we believe
to be a special point in the enlarged Kahler structure moduli space
of the Calabi-Yau model for the $r>>0$ limit.
We then
arrive to a Landau-Ginzburg orbifold model with the following form after
a rescaling of the fields
\be
L_{1}=\int d^2y d^4\theta\sum_i\bar{X}_iX_i
-\int d^2y d^2\theta \sum_iW^i(X).\label{lgo1}
\ee

In going through the point $r=0$, the size
of the Calabi-Yau shrinks to zero  leaving us with a singular
manifold.  However, the Kahler moduli space has as complex
coordinate $t=r+i\theta$.  Thus we may analytically continue
about $r=0$ by considering a path with $\theta\ne0$ and arrive at  the
Landau-Ginzburg phase without encountering any singularities \cite{can}.

It is suggestive the fact that the information that
remains in the Landau-Ginzburg
phase of (\ref{U}) is that of the tangent bundle.  On the other hand,
in the Calabi-Yau phase we find that the information that remains is
that of the hypersurface and the tangent bundle (through the mass term
for the fermions).  Thus if we construct two $(2,2)$ models
which have the same tangent bundle but with different embedding in the same
ambient space they will be dual because they will have the same Landau-Ginzburg
phase.  This situation is impossible to realize in the context of $(2,2)$
models.
The hypersurface embedded in the ambient space and its tangent bundle are in
one-to-one correspondence.  However, as we shall see, this is not the
case in $(0,2)$ models.  They allow different models to have the same
vector bundle data $F_{ij}$.

\section{Construction of $(0,2)$ Models}
\subsection{ The (0,2) Multiplets}
As noted in \cite{d1}, one may construct $(0,2)$ models which flow from a
Calabi-Yau phase to a Landau-Ginzburg phase by extending in a natural manner
the work of \cite{w1}.  For this we must introduce a set of chiral superfields
$X_i$.  In components,
they are  written, following the convention of \cite{w1}, as
\be
X=x+\sqrt{2} \theta^+\psi_+-i\theta^+\bar{\theta}^+(D_0+D_1)x,\non\\
\ee
It has the property
\be
\bar{{\cal D}}_+X=0,
\ee
where $
\bar{{\cal D}}_+$ is the gauge covariant derivative in superspace which
satisfies
\be
\bar{{\cal D}}_+=\frac{\partial}{\partial \theta^+}-i
\bar{\theta}^+(\partial_0+\partial_1+i(a_0+a_1)).
\ee
Of course, the fields $a_i$ belong to a $U(1)$ vector multiplet ${ A}$
which has
the following  expansion
\be
{ A}=a_0-a_1-2i\theta^+\bar{\lambda}_-
-2i\bar{\theta}^+{\lambda}_-+2\theta^+\bar{\theta}^+D,
\ee
in the Wess-Zumino gauge.

We must also make use of Fermi superfields $\Gamma_b$ whose components
are
\be
\Gamma_b=\gamma_b-\sqrt{2}\theta^+l_b-i\theta^+\bar{\theta}^+(D_0+D_1)\gamma_b.
\ee
These Fermi superfields will have the property
\be
\bar{{\cal D}}_+\Gamma_b=0.
\ee
\subsection{(0,2) Models which are not Deformations of (2,2) Compatifications}
We may now write the following action
\ba
L&=&L_{gauge}+L_{chiral}+L_{Fermi}+L_{D,\theta}+L_{W}\non\\
L_{gauge}&=&\frac{1}{8}\int d^2y d\theta^+d\bar{\theta}^+
\bar{\cal A}{\cal A}\non\\
L_{chiral}&=&\frac{-i}{2}\int d^2y d^2\theta(\bar{X}_i({\cal D}_0-{\cal
D}_1)X_i
+\bar{P}_j({\cal D}_0-{\cal D}_1)P_j)
\non\\
L_{Fermi}&=&-\frac{1}{2}\int d^2yd^2\theta \bar{\Gamma}_b\Gamma_b\non\\
\non\\
L_{D,\theta}&=&\frac{i t}{2}\int d^2y d\theta^+{\cal A}|_{\bar{\theta}^+=0}
+h.c.,
\non\\
L_{W}&=&\frac{-1}{\sqrt{2}}\int d^2y d\theta^+ \Gamma_bJ^b
|_{\bar{\theta}^+=0}+h.c.\label{SU2}
\ea
where ${\cal A}=[\bar{{\cal D}}_+,\partial_0-\partial_1+i{A}]$.
After performing the supercoordinate integration we arrive  at
the following bosonic potential \cite{w1}
\be
U(x_i)=\frac{1}{2}(\sum_i q_i |x_i|^2+\sum_a q_{j}|p_j|^2-r
)^2 +\sum_b|J_b|^2 .
\ee
The first term is proportional to $D^2$, where $D$ is the auxiliary field of
the vector multiplet.

We will consider the case in which
$J_b=\sum_j P_j F_{ji},\ b=i=1,...,n$ and $J_b=W_j,\ b=j=n,...,2n-4$,
and $\Gamma_b=\Lambda_i,\ b=i=1,...,n$ and $\Gamma_b=\Sigma_j,\
b=j=n,...,2n-4$,
and denote the charges of the $\Lambda_i$ by $n_i$ and the charges of the
$\Sigma_j$'s by $-d_j$.

The bosonic potential of the model will then be
\be
U(x,p)=\sum_j|W_j|^2
+\sum_i|\sum_jp^jF^i_j|
+\frac{1}{2}(\sum_i q_i|x_i|^2+q_{j}|p_j|^2-r)^2
\label{uf}
\ee
Let us then study  this potential in the limit in which $r>>1$ for the
case in which all the charges $q_i$ are positive and all the charges
$q_j$ are negative.  The vanishing
of the D-term, equivalently the vanishing of the last term in (\ref{uf}),
implies for this
case, that the fields $x_i$
cannot vanish simultaneously.
Since the $F^i_j$ vanish simultaneously only at the origin
it implies that all the $p_j$'s must vanish.  As an end result, we arrive at a
 $\sigma$ model which has a target space given by the locii of
the hypersurfaces $W_j$ embedded in ${\bf WCP}_{q_1,...,q_n}$.
In order for this manifold $M$ to be a Calabi-Yau manifold,
it must satisfy
\be
\sum_{i}q_i-\sum_{j} d_j=0.\label{cy}
\ee
where $d_j$ are the degree of the homogeneous polynomials $W_j$ and
$q_i$ are the charges of the chiral fields $X_i$ with respect to the
gauge field.
The natural question which arises is what role do the polynomials $F_j^i$
play.  For this we must study the left moving massless fermions $\lambda_i$'s.
As shown
in  \cite{d0}, the fermions which couple to these polynomials
transform as
sections of a holomorphic vector bundle $E$ over the manifold $M$ on which
the string propagates.  As all the vector bundle data $F_{ij}$ cannot vanish
simultaneously, we find that $j$ linear combinations of the $\lambda_i$'s
pick up a mass through the term in the Lagrangian
\be
\bar{\psi}_{p_{j}}\lambda_iF^{i,j}
\ee
where $\bar{\psi}_{p_j}$ is the fermionic component of the chiral multiplet
$P_j$.

The linear combinations which pick up a mass can be
read off the sequence
\be
0\to E\to\oplus^{r+1}_{a=1}{\cal O}(n_a)
\to^{\otimes F_{ij}(x)}{\cal O}(m)\to 0,
\ee
with
$\sum_j |q_{j}|=\sum_i n_i$.
Thus, we see that the polynomials $F_{ij}$ define the vector bundle $E$ over
$M$.
If we desire to construct a stable vector bundle which is not a deformation
of $TM$,
there are certain conditions we must impose.

The first condition demands that
the vector bundle $E\rightarrow M$ must yield an anomaly free theory.
This implies \cite{d0} that the second Chern class, $c_2(E)$, must be
equivalent
to the second Chern class of the tangent bundle $TM$.
  Then, in order to have
spinors defined over this vector bundle, the first Chern class of $E$
must be a multiple of two\footnote{$c_1(E)$ must also be positive otherwise $E$
is
never stable}.  We will take this class to vanish because it will also
guarantee that the Donaldson-Uhlenbeck-Yau condition \cite{d0} is satisfied,
a condition needed for $E$ to exist.
These two conditions impose respectively
the  constraints
\ba
c_2(E)&=&c_2(T)\\
c_1(E)&=&0.
\label{cond}
\ea
These may be also formulated in the form
\ba
\sum_j q_{j}^2-\sum_i n_i^2&=&
\sum_{j}{d^2_j}-\sum_{j}{q_j^2}\\
\sum_k |q_{j}|-\sum_i n_i
&=&0.
\label{conda}
\ea

The first condition implies that two different Calabi-Yau manifolds which
have the same vector bundle data could be dual as we shall soon show.

In the limit in which $r>>0$ we find a Calabi-Yau phase in which the
left moving
fermions which couple to the polynomials $F_{ij}$ transform as sections
of the vector bundle $E$.

At $r<<0$, in the LGO phase, we encounter an effective
superpotential given by
\be
W_{eff}=\sum_j\Sigma_j W^j+\sum_a P^jF_{ij}\Lambda^i\label{weff2}
\ee
This potential differs in the first term
from the potential encountered in the (2,2) models.  It is this term
which does not allow us to formulate the Higgs mechanism in
a straightforward manner.  Rather we should  find a way to make the
fermions $\Sigma_j$ massive.

\subsection{ (0,2) Compactifications As Deformations Of (2,2)
Compactifications}

Another possibility which we will use are deformations of (2,2) models.
These models  have also been treated in \cite{w1}.  They have in addition
to the fields used in the previous (0,2) model a fermi multiplet which
contains the field that was the scalar superpartner $\sigma$
of the gauge field $A$ in
the (2,2) model.  This field will appear with its fermionic partner
$\bar{\beta}$
in the
supermultiplet $\epsilon$.
In addition,  the fermi fields $\Gamma_b$ will have a different expansion
than those previously used in the above $(0,2)$ model.  They will satisfy
\be
\bar{{\cal D}}_+\Gamma_b=q_b\epsilon\phi_b.
\ee
where $\phi_b$ is a chiral field $X_i$ or $P_j$ and $q_b$ the charge of
the chiral field.
With these modifications we arrive at the following bosonic potential
\ba
U(x,p)&=&\sum_j|W_j|^2
+\sum_i|\sum_jp_jF^j_i|^2
+\frac{1}{2}(\sum_i q_i|x_i|^2+q_{j}|p_j|^2-r)^2\non\\
&& +|\sigma|^2(\sum_i |q_i|^2|x_i|^2+\sum_j |q_j|^2|p_j|^2).
\label{uf33}
\ea
We see that the last term in (\ref{uf33}) is the one which allows us to claim
that this model will yield a (2,2) deformation provided that we introduce
a fermionic gauge symmetry associated to the multiplet $\epsilon$.
This can only be done if
\be
\sum_i X_iF^i_j=d_j W_j
\ee
holds.

Let us then, study  this potential in the limit in which $r>>1$.  In this
limit, we find that the fermions $\Lambda_i$ transform as sections
of the deformed tangent bundle over a Calabi-Yau manifold defined
by the vanishing of the $W_j$'s on ${\bf WCP}_{q_1,...,q_n}^{n-1}. $
The vector
bundle over this manifold will be stable. As in the previous subsection
$j$ linear combinations of the $\Lambda_i$'s will pick up a mass
through  the mass term
\be
\bar{\psi}_{p_j}\lambda_iF^{ij}.
\ee
But another linear combination will also pick up a mass though the term
\be
\bar{\beta}\lambda_i\bar{\phi}^i
\ee
due to the presence of the multiplet $\epsilon$.

Thus, given two pairs $(M,E)$ where $M$ is a manifold and $E$ is a vector
bundle, with the same vector bundle data $F^{ij}$ we see that if one is a
(2,2) deformation it will have a rank 3 ($E_6$ gauge group) while  if the
second is not a (2,2) deformation it will have a stable rank 4 ($SO(10)$ gauge
group).  Can such pair be found to be dual?  The answer is yes as we shall
soon see.

  In the
$r<<0$ limit we find a HLO theory whose superpotential is
\be
\sum_{i,j} P^jF_{ji}\Lambda^i.
\ee
In this case, the fermions $\Sigma_j$ have picked up a mass
with the fermion component of the multiplet $\epsilon$ which contains the
scalar partner of the gauge field.

\subsection{ Higgs (0,2) Compactifications}

Another possibility is to add additional gauge fields,
chiral fields and fermi
fields to the lagrangian (\ref{SU2}). The field content of the model is
summarized below.

Our old gauge field
${A}$ will be acompanied by other gauge fields  ${ B}_j$.
Then, the charge of the fields  for the case in which we have
two hypersurfaces of degree $d_j, j=1,2$
are given by the array
$(q_A ;q_{B_1} ,q_{B_2})$.

The field content and charges are as follows.

Chiral primaries $X_i$ with charges $(q_i;0,0)$. These are the
fields which are used to define the CY manifold and they are present in
(0,2) models.
Chiral primaries $P_j$ with charges $(q_{j};0,0)$.
$P_j$ are  also present in all (0,2) models.
Chiral primaries $G_j$ with charges $(e_j;0,0)$.
These fields are not present in the previous (0,2) models. The charges $e_j$
will be fixed by demanding gauge invariance of the superpotential with respect
to the gauge field $A$.
Chiral primaries $Y_j$ with charges
$(b_{j_1};\beta_{j_1},0)$,
$(b_{j_2};0,\beta_{j_2})$.
These fields are not present in
the previous (0,2) models.
Chiral primaries $S_j$ with charges
$((-b_{j_1}-2e_{j_1})/2;-\beta_{j_1}/2,0)$,
$((-b_{j_2}-2e_{j_2})/2;0,-\beta_{j_2}/2)$.
These fields are not present in
the previous (0,2) models.
Fermi fields $\Lambda_i$ with charges $(n_i;0,0)$.  Some of these fields will
remain massless in the CY phase and will define the vector bundle.
These fields are present in
the previous (0,2) models.
Fermi fields $\Sigma_j$ with charges $(-d_j;0,0)$.
These fields will remain massive in the LGO phase.  They are present
in the previous model but are massless in the LGO phase and couple to the
hypersurface polynomials.
Fermi field $\Xi_j$ with charges
  $(-d_{j_1}-e_{j_1};-\beta_{j_1},0)$,
  $(-d_{j_2}-e_{j_2};0,-\beta_{j_2})$.
These fields will remain massive in the HLG phase.  They are not present
in the previous (0,2) models.
Fermi fields $\Upsilon_j$ with charges
$((-b_{j_1}+2e_{j_1})/2;-\beta_{j_1}/2,0)$,
$((-b_{j_2}+2e_{j_2})/2;0,-\beta_{j_2}/2)$.
The additional fields we have introduced
will remain massive in the HGO and Calabi-Yau  phases.  They are not present
in the previous (0,2) models.
With these choices of charges we find that the cancelation of all the
anomalies reduces
to the ones previously encountered in other (0,2) models, $c_2(E)=c_2(TM)$.
Since the charges of the additional fields are
functions of $b_j$ and $\beta_j$
which are arbitrary constants, the charges are quite arbitrary,
although it is required that the central charge of the theory in the HLO phase
as well as the Calabi-Yau phase be
consistent with the compactification.

The complete action is
\ba
L&=&L_{kinetic}+L_{D,\theta}+L_{W}\non\\
L_{D,\theta}&=&\frac{i t}{2}\int d^2y d\theta^+({\cal A}|_{\bar{\theta}^+=0}
-\sum_j{\cal B}_j)|_{\bar{\theta}^+=0})+h.c.\non\\
L_{W}&=&\int d^2y d\theta^+ (\sum_{ij} \Lambda_i P_jF^{ij}+\sum_j\Sigma_j
W_j\non\\&&
+\sum_j(\Sigma^j M_{jk}(P) G^k+\Xi_j Y_j G_j+\Upsilon_jS_j Y_j))
|_{\bar{\theta}^+=0}+h.c.\label{SU1}
\ea

The  matrix $M(P)$ has entries which are
holomorphic functions of $P_j$ with appropiate
charges to guarantee gauge invariance of the superpotential.
Its determinant does not vanish anywhere over ${\bf WCP}_{q_{j_1},...}^3$.
For
appropriate values of $b_j$ and $\beta_j$,
in the limit $r>>0$
we find that the fields, which were not present in the previously
studied (0,2) models, become massive.  In addition, all chiral primaries have
vanishing expectation values with the exception of $Y_j$, and thus
decouple from the remaining fields.  The effective action for the
massless fields is a non linear $\sigma$ model whose vector bundle
is determined by the polynomials $F_{ij}$ only, as was the case in the
previous subsection, since there $\lambda_i$'s cannot couple to
the $\epsilon$ multiplet which is absent from this model.  Thus,
the bundle defining data $F_{ij}$ will give rise to a rank 3 gauge bundle
for (2,2) models but the same data in the model presented in this section
will define a rank 4 vector bundle as was the case in the previous subsection.

On the other hand,
for $r<<0$ we find that the fields $\Sigma_j$ become massive and
we arrive at  an HLG phase for the massless fields whose superpotential
is given by

\be
L=\sum_{ij} \Lambda_j P_i F^{ij}.
\ee
The fermi fields $\Xi_j$ remain massless and their charges
are quite arbitrary.  In fact, their charges can vanish.

\section{Higgs Compactifications: $E_6-SO(10)$ Case}

The best example to study $E_6$ breaking to $SO(10)$ is the quintic.
To exhibit this breaking  we must consider a linear $\sigma$ model used
in subsection 3.4.  Since we will consider the quintic we will have $j=1$
only.  Thus, we will drop this index.

The complete action is
\ba
L&=&L_{kinetic}+L_{D,\theta}+L_{W}\non\\
L_{D,\theta}&=&\frac{i t}{2}\int d^2y d\theta^+({\cal A}|_{\bar{\theta}^+=0}
-\sum_j{\cal B})|_{\bar{\theta}^+=0})+h.c.\non\\
L_{W}&=&\int d^2y d\theta^+ (\sum_{i}
\Lambda_i P F^{i}+\sum\Sigma W\non\\&&
+(\Sigma  P  G+\Xi Y G+\Upsilon S Y))
|_{\bar{\theta}^+=0}+h.c.\label{Su1}
\ea

The bosonic potential reads
\ba
U&=&|W+ p g|^2+|p|^2\sum_{i}|F_{i}|^2+|y|^2|g|^2 +|y|^2|s|^2\non\\
&&+(\sum_iq_i|x_i|^2+e|g|^2+b|y|^2
-(b+2e)/2|s|^2+q_p|p|^2-r)^2\non\\
&&+(-\beta/2 |s|^2+\beta|y|^2-r)^2.\label{ue6}
\ea
The polynomial $W$ is homogeneous of degree $5$, and $x_i F^i=W$.
The charge $q_p=-5$ is negative and the charges $q_i=1$ are all positive.
The charges of the $\Lambda_i$'s are  equal to the charges of the $X_i$'s.
The charge of $\Sigma$ is $-5$ as required by gauge invariance.
The charge $e$ is determined by demanding gauge invariance of the
superpotential.  With these choices of charges
the gauge anomalies for the $A$ and $B$
fields, as well as the mixed anomaly, cancel.   However, we have not set
the charge of the field $\Xi$.  We will take the charge of this field
to vanish.  This will fix the charges of $Y$ and $S$.

For large and positive $r$ we see that $y$ cannot vanish and thus
$s$ vanishes.  Similarly, $g$ must vanish.  For very large $\beta$ we
find that the $x_i$ cannot vanish simultneously
 and given the fact that the $F_i$
vanish simultaneously only at the origin, $p$ must vanish.  The end result
is a hypersurface, given by the vanishing of $W$, embedded in ${\bf CP}^4$.
Given the degree of $W$ we find a manifold with vanishing Ricci tensor.
The ambient space of the manifold is determined by the vanishing of the
second line in (\ref{ue6}) after moding out by the $U(1)$ symmetry
introduced by the gauge field $A$.  The $U(1)$ symmetry introduced by the
gauge field $B$ fixes the phase of the field $y$.  One linear
combination of the $\lambda_i$'s
picks up a mass with the fermionic component of the $P$ multiplet.
The left moving  fermions then
transform as sections of the vector bundle determined by the $F_i$.  Given
our choice of data, the vector bundle will be an extension of the tangent
bundle $E=T\oplus {\cal O}$.  It has rank 4 and
the gauge group associated to it is $SO(10)$.
The anomaly cancellation reduces to
$$c_2(E)=c_2(T)$$
which is the standard form of the anomaly cancellation for a vector
bundle over a Calabi-Yau manifold.  The fields
 $g$, $s$, $y$, $\xi$, $\upsilon$
are all massive in this phase. The central charge of this phase is \cite{d0}
$(10,9)$.
This model is in agreement with \cite{ds}
where the existence of
$E_6$ breaking directions in (0,2) compactifications where found for the
quintic.

For negative and large values of $r$, we find that $s$ cannot vanish.  Which
implies that $y$ vanishes.  These two fields along with $\upsilon$ become
massive.  The gauge symmetry of $B$ is used to set the phase of $s$.
The vanishing of the D-term of the field $A$ implies that $p$ does not vanish.
By transversality of the $F_i$'s all $x_i$ vanish.  This in turn forces
$g$ to vanish.  The fermions $\Lambda_i$ remain massless while $\sigma$
picks a mass with the fermionic component of $G$.  Perhaps the most important
fermion is $\xi$.  This fellow remains free, massless and it is uncharged.
The effective superpotential for the Landau-Ginzburg
action found in this phase is
\be
\int \Lambda_i F^i.
\ee
This action is nothing but the effective superpotential for the $E_6$
Landau-Ginzburg phase of the quintic.  It has central charge (9,9).  However,
since there is also a free complex left moving fermion, the total charge
of the model is $(10,9)$ as it was in the Calabi-Yau phase.  The complex free
fermion will join the remaining  $8$ real free fermions. In all we will
have in this phase $10$ free fermions which together with the
left moving
$U(1)$ generate $E_6$.  Thus we have continuously gone from the
Calabi-Yau phase of a quintic
with $SO(10)$ gauge group to the Landau-Ginzburg
quintic with $E_6$ gauge group. This in turn is connected in a continuous
manner
to the Calabi-Yau phase of the same quintic \cite{w2}.  It has recently
been shown that the phases with $E_6$ gauge group are stable against
world sheet instanton effects \cite{sw}.
As we have gone in a continuous manner from
the $E_6$ quintic to the $SO(10)$ quintic we may expect no additional
instanton contributions which may destabilize the $SO(10)$ phase.  We thus
expect to have a stable $SO(10)$ phase for the quintic.

This is in agreement with the statements made in \cite{d1} where it was
argued that the $SO(10)$ quintic in its LGO phase was unstable.
This is because
the argument of \cite{d1} was phrased at the $SO(10)$ LGO point where
there is a quantum symmetry associated to the discrete group which
orbifolds the LG theory.  This symmetry which prevents gauginos of picking
a mass at the LGO
point  is no longer present at the $SO(10)$ Calabi-Yau phase which
is  continuosly connected to the $E_6$ LGO phase.  Furthermore, from
the above analysis it follows that the mass of the massless states depends
on the Kahler moduli.  This is not the case in (2,2) compactifications which
have $E_6$ gauge group.  However, (2,2) compactifications have an $N=2$
 special
geometry which prevents the massless states from picking up a mass.  In (0,2)
compatifications we find that the special geometry is absent and that there
are additional moduli: gauge moduli.  These moduli, can in principle,
mediate between the
Kahler moduli and the massless states
in non trivial ways to make some of these massive \cite{ds}.

\section{Higgs Compactifications: $E_6-SU(5)$ Case}
The simplest examples to consider are models in which the gauge
group $E_6$ breaks down to $SU(5)$ without changing the number of
effective generations are along the lines of the quintic.
However, we must use complete intersection Calabi-Yau  manifold
and use two more
additional gauge fields $B_1$ and $B_2$.  The action is
\ba
L&=&L_{kinetic}+L_{D,\theta}+L_{W}\non\\
L_{D,\theta}&=&\frac{i t}{2}\int d^2y d\theta^+({\cal A}|_{\bar{\theta}^+=0}
-\sum_j{\cal B}_j)|_{\bar{\theta}^+=0})+h.c.\non\\
L_{W}&=&\int d^2y d\theta^+ (\sum_{ij}
\Lambda_i P_jF^{ij}+\sum_j\Sigma_j W_j\non\\&&
+\sum_j(\Sigma^j M_{jk}(P) G^k+\Xi_j Y_j G_j+\Upsilon_jS_j Y_j))
|_{\bar{\theta}^+=0}+h.c.\label{Su5}
\ea

We will take the charges of the $x_i\, i=1,...,6$ to be unity.
The charges of
$P_j\, j=1,2$ will be $-3$ for both.  The homogeneous polynomials $W_j$ will
be of degree $3$, thus fixing the charges of the fermi fields $\Sigma_j$.
The charges of the fermi fields $\Xi_j$ will vanish.  The polynomials
$F_{ij}$ will be of degree $2$ and will satisfy
$$x_iF^{ij}=W^j.$$

For large and positive $r$ we find a three complex dimension
Calabi-Yau phase given by the embedding of $W_j$ in ${\bf CP}^5$
with a gauge group $SU(5)$.  The central charge of this model
is (11,9). The five massless left moving
fermions $\lambda_i$'s which did not pick up a mass through
the mass term $$\lambda_i\bar{\psi}_{p_j}F^{ij}$$ transform as sections
of the vector bundle which is an extension of the tangent bundle
$$E=T\oplus{\cal O}\oplus{\cal O}.$$
The rank of it will be five and the gauge group associated to it
will then be $SU(5)$.
 All the fields which are charged
with respect to the $B_j$ gauge fields are massive. The $G_j$'s are
also massive.

For large and negative
values of $r$ we  find a Hybrid Landau-Ginzburg phase (HLG).  The mechanism
to make the fermions $\Sigma_j$ massive is present provided the holomorphic
mass matrix $M(p)$ has a  nonvanishing determinant.  These fermions
are replaced by the
chargeless fermions $\xi_j$ which become massless in this phase.
The superpotential for the action of the compactification is
\be
\int\sum_j \Lambda_iF^{ij}.
\ee
It has charge (9,9).  With the complex $\Xi_j$ the central charge goes up to
(11,9) as in the Calabi-Yau phase.
It is the phase of the $E_6$ Hibrid Landau-Ginzburg.  It is continuously
connected to the $E_6$ Calabi-Yau phase given by the embedding of
$W_j$ in ${\bf CP}^5$.  As the free complex fermions $\Xi_j$ are uncharged
they may join the remaining free real fermions to construct an $E_6$
gauge group.  We then have, continuously connected a Calabi-Yau phase
with $SU(5)$ gauge group to an HLG phase with $E_6$ gauge group.  The latter is
continuously connected to the
Calabi-Yau phase with $E_6$ gauge group.   The same manifold but different
gauge groups and vector bundles are continuosly connected through the
mechanism presented here.

\section{Conclusion}

We have succeeded in formulating the Higgs mechanism in (0,2)
compactifications.
This was achieved by introducing additional fields to the ones which
usually make their appearence in the literature.  More work is needed to
determine the stability of the $SO(10)$ and $SU(5)$ Calabi-Yau phases.

\pagebreak

\end{document}